\documentclass[fleqn,twoside]{article}
\usepackage{amsmath,epsf}
\usepackage{espcrc2}

\usepackage{graphicx}
\usepackage[figuresright]{rotating}

\newcommand\beq{\begin{equation}}
\newcommand\eeq{\end{equation}}
\newcommand\bal{ \begin{align}}
\newcommand\eal{\end{align} }

\newcommand\eqn[1]{\label{eq:#1}} 
\newcommand\Eq[1]{Eq.~\eqref{eq:#1}} 
\newcommand\Eqs[2]{Eqs.~(\ref{eq:#1},\ref{eq:#2})} 
\newcommand\half{{\textstyle{\frac{1}{2}}}} 
\newcommand\fourth{{\textstyle{\frac{1}{4}}}}

\newcommand\bfx{\mathbf{X}}
\newcommand\bfy{\mathbf{Y}}

\newcommand\bfz{\mathbf{Z}}

\newcommand\bfXi{\boldsymbol{\Xi}}
\newcommand\bfU{\boldsymbol{\Upsilon}}

\newcommand{\CN}{{\cal N}}

\newcommand{\CQ}{{\cal Q}}

\newcommand{\CL}{{\cal L}}

\newcommand{\bfn}{{\bf n}}

\newcommand{\xh}{\mathbf{\hat{e}}_1}
\newcommand{\yh}{\mathbf{\hat{e}}_2}

\newcommand{\ih}{\mathbf{\hat{e}}_i}
\newcommand{\jh}{\mathbf{\hat{e}}_j}

\newcommand{\Tr}{{\rm Tr}\,}
\newcommand{\sla}[1]%
        {\kern .25em\raise.18ex\hbox{$/$}\kern-.55em #1}
\newcommand{\vev}[1]{\langle #1 \rangle}
\newcommand{\mybar}[1]%
        {\kern 0.6pt\overline{\kern -0.6pt#1\kern -0.6pt}\kern 0.6pt}
\newcommand{\dig}{\kern-1.5pt \raisebox{.9ex}{$\cdot$}  \kern1.5pt
  \raisebox{0ex}{${\mathbf\cdot}$}\kern1.5pt \raisebox{-.9ex}{$\cdot$}} 
\newcommand{\digb}{\kern-1.5pt \raisebox{.75ex}{$\cdot$}  \kern1.5pt
  \raisebox{0ex}{${\mathbf\cdot}$}\kern1.5pt \raisebox{-.75ex}{$\cdot$}} 
\newcommand{\digc}{\kern-1.5pt \raisebox{1.05ex}{$\cdot$}  \kern1.5pt
  \raisebox{0ex}{${\mathbf\cdot}$}\kern1.5pt \raisebox{-1.05ex}{$\cdot$}} 
\newcommand{\drawsquare}[2]{\hbox{%
\rule{#2pt}{#1pt}\hskip-#2pt
\rule{#1pt}{#2pt}\hskip-#1pt
\rule[#1pt]{#1pt}{#2pt}}\rule[#1pt]{#2pt}{#2pt}\hskip-#2pt
\rule{#2pt}{#1pt}}
\newcommand{\Yfund}{\,\raisebox{-.5pt}{\drawsquare{6.5}{0.4}}\,}
\newcommand{\Ybarfund}{\mybar{\raisebox{-.5pt}{\drawsquare{6.5}{0.4}}}\,}%

\title{Recent Developments in Lattice Supersymmetry}

\author{David B. Kaplan\address[INT]{Institute for Nuclear Theory \\
    Box 351550, University of Washington \\ Seattle, WA 98195-1550,
    USA}\thanks{This work was  supported in part by DOE grant DE-FGO3-00ER41132}}%

\begin{document}

\begin{abstract}
I discuss a new approach to constructing lattices for 
gauge theories with extended supersymmetry.  The
lattice theories themselves respect certain supersymmetries, which in
many cases allows the target theory to be obtained in the continuum
limit without fine-tuning.  
\vspace{1pc}
\end{abstract}

\maketitle

\section{Introduction}

Supersymmetric gauge theories are known to exhibit diverse fascinating
phenomena (See, for example,
\cite{Seiberg:1994aj,Terning:2003th,Klebanov:2000me}).  These include
phenomena seen in QCD, such as confinement 
and chiral symmetry breaking. 
In some cases one can demonstrate phenomena suspected to be important
in QCD, such as magnetic monopole condensation and instantons.  Others
exhibit phenomena which are decidedly unlike QCD, such as massless
composite fermions.  There
exist pairs of theories which have quite different Lagrangian
descriptions (such as different gauge groups) which are believed to be
dual to each other.  Some of these 
theories are expected to possess nontrivial conformal fixed points in
the infrared.  A particularly special case, $\CN=4$ supersymmetric
Yang-Mills 
theory (SYM) in four dimensions is thought to be exactly conformal and
to be self-dual.  Understanding these theories in detail would greatly
expand our knowledge of field theory.  These theories could be of more
than pedagogical importance however.  Many have speculated that
strongly coupled SYM theories could explain the hierarchy between the
weak and GUT scales and  the baffling pattern of quark and lepton
masses. Beyond that, string theory
and all of quantum gravity is thought to be related to such theories
in the large-$N_c$ limit, where $N_c$ is the number of colors of the
gauge group.

Up until recently there has not existed a non-perturbative regulator
for such theories, and so our knowledge of their behavior has been
limited to certain analytical calculations and informed speculation.

Recently there have been a number of interesting developments in
lattice supersymmetry, such as work on the $d=2$ dimensional
Wess-Zumino model
\cite{Catterall:2001fr,Catterall:2003ae,Beccaria:2001tk,Fujikawa:2002ic},
on $d=4$, 
$\CN=1$ SYM theory \cite{Montvay:2001aj,Feo:2002yi,Feo:2003km}, as
well as other supersymmetric theories
\cite{Catterall:2003wd,Nishimura:2003tf,Itoh:2002nq}. In this talk I
will tell you about  recent progress I have made 
with my collaborators A. Cohen, E. Katz, and M. \"Unsal in
constructing lattice theories whose continuum limits are various SYM
theories in various dimensions
\cite{Kaplan:2002wv,Kaplan:2002zs,Cohen:2003xe,Cohen:2003qw}.  In many
cases we can show that the 
target theory is obtained in the continuum limit without the need for
fine-tuning of couplings.  This is accomplished by maintaining some
unbroken supersymmetry at finite lattice spacing, an approach shared
by the recent work discussed at this conference by Simon Catterall
\cite{Catterall:2001fr,Catterall:2003wd}. The lattices that 
result are quite unlike ones that are usually considered by lattice
theorists. I intend to exhibit some of their peculiar and almost
magical properties, such as the emergence of chiral symmetries in the
continuum without having to resort to overlap or domain wall fermions.

\section{Accidental Supersymmetry}
\label{sec:2}

Formulating a  supersymmetric lattice gauge theory is difficult, since
supersymmetry is actually an extension of the Poincar\'e algebra,
which is explicitly broken by the lattice.  In particular, the central
feature of the super-Poincar\'e algebra is the anti-commutator of a
supercharge $Q_\alpha$ and its conjugate $\mybar Q_{\dot \alpha}$,
which yields the generator of infinitesimal translations,
$P_\mu$.  Schematically, 
$\{Q,\mybar Q\} \sim 2 P$.  On the lattice there are no infinitesimal
translations, and therefore the supersymmetry algebra must be broken.

Nevertheless, we are quite familiar with the continuum limit of a lattice
theory possessing more symmetry than the lattice action itself, and we might
hope to construct non-supersymmetric lattices with supersymmetric
continuum limits. Certainly it ought to be possible to accomplish with
enough fine-tuning of coupling constants, but that approach is neither
feasible in practice nor theoretically satisfying. An obvious paradigm
to emulate is the
emergence of Poincar\'e symmetry without fine-tuning in lattice QCD.  The way this works
is that the exact symmetries of the lattice theory --- specifically
gauge symmetry and the hypercubic crystal symmetry --- forbid
operators with dimension $\le 4$ which violate Poincar\'e symmetry.
Thus at weak coupling, the theory flows to a Poincar\'e symmetric
point in the infrared without any fine-tuning of couplings
required. In this sense, Poincar\'e symmetry emerges as an
``accidental'' symmetry in the continuum.  As we well know, one is not
always so lucky: witness the difficulties in obtaining chiral symmetry
in the continuum limit of lattice QCD.

Although we cannot construct lattices which obey the super-Poincar\'e
algebra, we may still hope to find lattices for which supersymmetry
emerges as an accidental symmetry in the infrared. 

A simple example of a non-supersymmetric theory with a supersymmetric
limit in the infrared is an $SU(N)$  gauge theory with a single Weyl spinor
transforming in the adjoint representation, respecting an exact $Z_{2N}$
chiral symmetry, the anomaly-free subgroup of phase rotations of the
fermion. Such a theory has $\CN=1$ SYM as its infrared 
limit, since the only possible supersymmetry violating relevant
operator allowed by the gauge 
and spacetime symmetries is a fermion mass, and that is forbidden by
the anomaly-free $Z_{2N}$ chiral symmetry \cite{Kaplan:1984sk}. It is possible to construct
such a theory on the lattice using 
overlap or domain wall fermions
\cite{Neuberger:1998bg,Kaplan:1999jn,Nishimura:1997vg,Fleming:2000fa}.
It is also possible to dispense with 
the chiral symmetry and to use Wilson fermions, making one fine-tuning
(setting the fermion mass to zero) in order to obtain the supersymmetric target
theory \cite{Montvay:2001aj}. This is a very interesting theory to study, but as it
requires dynamical fermions to exhibit supersymmetry, the technical
challenges to simulating the theory are great.

In four dimensions, the above example of pure  $\CN=1$ SYM theory is
the only supersymmetric theory without scalar fields. If one wishes to
simulate any other supersymmetric theory in $d=4$ dimensions, then
scalars must be introduced, and the situation looks grim.  That is
because among the plethora of relevant supersymmetry violating
operators that must now be considered, one is a mass term for the
scalar. There are only two symmetries which can
forbid a scalar mass term.  The first is a shift symmetry,
$\phi \to \phi+f$, where $f$ is a constant.  This results in $\phi$
only having derivative interactions...it is a Goldstone boson.
Such a theory cannot describe most supersymmetric theories of
interest, including SYM theories, in which scalar fields have
gauge  interactions.

The second symmetry which can forbid mass terms is supersymmetry. This
reasoning seems to have led us in full circle: due to the difficulties
in realizing supersymmetry exactly on the lattice, we are led to look for
non-supersymmetric theories exhibiting accidental supersymmetry; but
accidental supersymmetry seems to require exact supersymmetry in order to forbid scalar
mass terms.  

All that remains to us is a sort of compromise:  perhaps
we can construct a lattice with a little bit of exact supersymmetry,
enough to forbid relevant operators which violate any of the target
theories more extensive supersymmetry.  There are lots of reasons to
expect this approach to fail for SYM theories, however:

\smallskip
\noindent $\bullet$\  
Presumably the exact supersymmetry would require scalar, fermions and
  gauge fields to exist in the same multiplet.  But then if gauge
  fields live on links, one would expect their supersymmetric
  partners to as well.  But how can
 spin zero bosons reside on links, which would require them
 to transform nontrivial under the exact lattice
  rotations, and hence (one would expect) under continuum Lorentz
  transformations?

\smallskip\noindent $\bullet$\  
With so many particles of different spin around, it would seem
  difficult  to even realize accidental Lorentz
  symmetry in the continuum due to the large number of relevant Lorentz
  violating interactions.  Lorentz violation has been the outcome of
  many attempts to construct lattices with partial supersymmetry, such as in
  ref.~\cite{Banks:1982ut};

\smallskip\noindent $\bullet$\  
The SYM target theories typically have large chiral symmetries
  (called $R$-symmetries)
  which do not commute with the supersymmetry. For example, $\CN=4$
  supersymmetry in $d=4$ has an $SU(4)$ $R$-symmetry, under which the fermions
  transform as the four dimensional defining representation, while the
  scalars transform as the six dimensional antisymmetric tensor. Our
  experience with lattice QCD suggests that we would either have to
  fine-tune the theory to realize these symmetries, or else employ
  overlap or domain wall fermions.  In the latter cases it is hard to
  imagine how there could be a supersymmetry relating such fermions to
  scalars or gauge bosons, which have quite different implementations.

\smallskip

Happily, we have found that there exist lattices which get around all
of the above objections. 
 The technical details of how to construct these lattices are
given in
refs.~\cite{Kaplan:2002wv,Kaplan:2002zs,Cohen:2003xe,Cohen:2003qw};
here I will focus instead on what the lattices 
look like, and how they work.  I will begin by discussing a lattice theory
for SYM in two Euclidean dimensions, possessing four real
supercharges.

(A matter of nomenclature:  rather than designating the amount of
supersymmetry by $\CN=1$,
$\CN=2$, etc. in dimensions $d<4$,  which gets confusing, I will instead refer to the number of real supercharges $\CQ$.
For comparison,  $\CN=1$, $\CN=2$ and $\CN=4$ SYM theories in  $d=4$ possess
$\CQ=4$, $\CQ=8$ and $\CQ=16$ real supercharges respectively.)

\section{Super Yang Mills in $d=2$ with $\CQ=4$}
\label{sec:3}

 This target theory is just what one gets upon
reducing conventional $\CN=1$ SYM theory in $d=4$ down to $d=2$.  The
particle content is  a two component gauge field $v_m$, one complex
scalar $s$, and a Dirac fermion $\psi$.  All of these fields transform
as adjoints under the gauge symmetry.  The Lagrangean in Euclidean
space is given by
\begin{equation}
\begin{aligned}
  \CL &=  \frac{1}{g_2^2}  \, \Tr
  \Biggl(\bigl\vert D_m s\bigr\vert^2 + \mybar \psi \,  D_m
  \gamma_m 
  \psi + \fourth v_{mn} v_{mn} \\ & +\sqrt{2}(\mybar \psi_L [ s, \psi_R]
  +\mybar\psi_R [ s^\dagger, \psi_L]) - \half
  [s^\dagger,s\,]^2\Biggr) 
  \eqn{targ2}\end{aligned}
\end{equation}
In the above equation, $v_{mn}$ is the gauge field strength, and $g_2$
is the gauge coupling.  Note that this theory exhibits an anomalous chiral $U(1)$
symmetry,
involving both $\psi$ and $s$.

\begin{figure}[tb]
\centerline{\epsfxsize=2.5 in \epsfbox{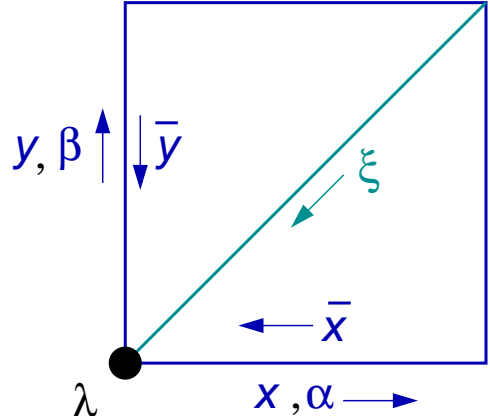}}
\caption{The lattice for SYM in $d=2$ with $\CQ=4$ supercharges. Latin
letters correspond to one-component complex bosons, Greek letters to
one component fermions.}
\label{fig:2dq4}
\end{figure}
Our lattice for this theory has the structure shown in Fig.~\ref{fig:2dq4}.
It consists of two complex bosons $x$ and $y$, and four one-component
Grassmann variables $\alpha$, $\beta$, $\lambda$ and $\xi$.  All
fields are $k\times k$ matrices  At each
site there is an independent $U(k)$ symmetry, which will become a
$U(k)$ gauge symmetry in the continuum;  $\lambda$ transforms
as a $U(k)$ adjoint, while the link variables are bifundamentals
$(\Yfund,\Ybarfund)$ or the conjugate, depending on the orientation of
the  arrows shown in
the figure.

\subsection{The lattice action and its classical limit}
\label{sec:3a}

The lattice action is given by

\begin{equation}\begin{aligned}
  \frac{1}{g^2} \sum_{\mathbf n} \Tr\Bigl[&\half \left(\mybar
    x_{\mathbf{n}-\xh} x_{\mathbf{n}-\xh} - x_{\mathbf n}\mybar
    x_{\mathbf n}\right.\Bigr.\\ 
&+ \left. \mybar y_{\mathbf{n}-\yh}
    y_{\mathbf{n}-\yh} - y_{\mathbf n}\mybar y_{\mathbf
      n}\right)^2
\\ & 
+2\,\bigl| x_{\mathbf n} y_{\mathbf{n} +\xh} -
  y_{\mathbf n} x_{\mathbf{n} + \yh}\bigr|^2
  \\ \Bigl.&
+\sqrt{2}\, \left(\Delta_{\bfn}(\lambda,\mybar x,\alpha)
    +\Delta_{\bfn}(\lambda,\mybar y,\beta) \right.\\&\left. 
- \Delta_{\bfn}(\xi,y,\alpha) + \Delta_{\bfn}(\xi,x,\beta)\right)\Bigr]
\end{aligned}
\eqn{lat2d}\end{equation}
where ${\mathbf n}$ denotes the lattice position, $\xh$ and $\yh$ are
unit lattice vectors, and
\begin{equation}\begin{aligned}
\Delta_{\bfn}(\lambda,\mybar x,\alpha) &=
\lambda_\bfn\left(\mybar x_{\bfn-\xh} \alpha_{\bfn-\xh} - \alpha_\bfn \mybar
  x_\bfn\right)\ ,\\
\Delta_{\bfn}(\lambda,\mybar y,\beta) &=
\lambda_\bfn\left(\mybar y_{\bfn-\yh} \beta_{\bfn-\yh} - \beta_\bfn \mybar
  y_\bfn\right)\ ,\\
\Delta_{\bfn}(\xi,y,\alpha) &= \xi_\bfn\left(y_\bfn\alpha_{\bfn+\yh}
    -\alpha_\bfn y_{\bfn+\xh}\right) ,\\
\Delta_{\bfn}(\xi,x,\beta) &= \xi_\bfn\left(x_\bfn\beta_{\bfn+\yh}
    -\beta_\bfn x_{\bfn+\yh}\right)
\end{aligned}
\eqn{delta}\end{equation}
(In fact I will need to add some more terms to the above action, as
discussed below in \S\ref{sec:3c}.) As it stands, this ``lattice'' action looks bizarre:
 there are no hopping terms and no lattice spacing defined. 
The action has ``classical flat directions'', or moduli.   We will choose to
expand the action about one particular point in this moduli space,
namely
\begin{equation}
\vev{x}=\vev{\mybar x}=\vev{y}=\vev{\mybar y} = \frac{1}{\sqrt{2}\,
{a}}\, \times {\boldsymbol {1_k}}\ .
\eqn{vevs}\end{equation}
Here $\boldsymbol {1_k}$ is the $k$-dimensional unit matrix, and ${a}$
will soon be interpreted as  the lattice spacing.  

First,  consider the bosonic part of the action. I rewrite the bosonic
variables as
\beq
x \equiv  \frac{1}{\sqrt{2}\, a} + \frac{s_1 + i v_1}{\sqrt{2}}\ ,\quad
y \equiv  \frac{1}{\sqrt{2}\, a} + \frac{s_2 + i v_2}{\sqrt{2}}\ 
\eeq
and expand in powers of $a$. For example, $\mybar
    x_{\mathbf{n}-\xh} x_{\mathbf{n}-\xh} - x_{\mathbf n}\mybar
    x_{\mathbf n}$ is expanded as
\begin{equation}
\begin{aligned}
&\left(\frac{1}{\sqrt{2}\, a} + \frac{s_1-iv_1}{\sqrt{2} }-
    a\frac{\partial\ }{\partial x} \frac{s_1-iv_1}{\sqrt{2}} +
      O(a^2)\right)\cr
\times &
\left(\frac{1}{\sqrt{2}\, a} + \frac{s_1+iv_1}{\sqrt{2}} -
    a\frac{\partial\ }{\partial x} \frac{s_1+iv_1}{\sqrt{2}} +
      O(a^2)\right)\cr
-&\left(\frac{1}{\sqrt{2}\, a} +
  \frac{s_1+iv_1}{\sqrt{2}}\right)\times\left(\frac{1}{\sqrt{2}\, a} +
  \frac{s_1-iv_1}{\sqrt{2}}\right)\cr
=& -i[v_1,s_1]-\frac{\partial s_1 }{\partial x} + O(a)\cr
=&-D_1 s_1 + O(a)\ ,
\end{aligned}
\nonumber\end{equation}
with $D_m = \partial_m + i [v_m,\cdot]$.  Note that while naively the
expression $(\mybar
    x_{\mathbf{n}-\xh} x_{\mathbf{n}-\xh} - x_{\mathbf n}\mybar
    x_{\mathbf n})$ looks to be $O(a^{-2})$, it is in fact  finite
as $a\to 0$ due to its commutator-like structure.

Carrying out this expansion in powers of $a$ of the various terms in
the lattice action \Eq{lat2d} I find
\beq
\begin{aligned}
& \frac{1}{g^2} \sum_{\mathbf n} \Tr\Bigl[\half \left(\mybar
    x_{\mathbf{n}-\xh} x_{\mathbf{n}-\xh} - x_{\mathbf n}\mybar
    x_{\mathbf n}\right.\cr
&\qquad\qquad+ \left.\mybar y_{\mathbf{n}-\yh}
    y_{\mathbf{n}-\yh} - y_{\mathbf n}\mybar y_{\mathbf
      n}\right)^2\Bigr]\cr
& =\ \ 
\frac{1}{g^2 a^2} \int d^2z \frac{1}{2}\Tr\left(D_1 s_1 + D_2
  s_2\right)^2 + O(a)\ ,
\end{aligned}
\eqn{terma}\eeq
and
\beq
\begin{aligned}
  &\quad\frac{1}{g^2} \sum_{\mathbf n}  2\,\Tr\bigl| x_{\mathbf n} y_{\mathbf{n} +\xh} -
  y_{\mathbf n} x_{\mathbf{n} + \yh}\bigr|^2 \cr
&=\ \ 
\frac{1}{g^2 a^2} \int d^2z\  \frac{1}{2}\left(D_1 s_2 - D_2
    s_1\right)^2 +\frac{1}{4} [s_1,s_2]^2\cr
&
\qquad\qquad\qquad\qquad+\frac{1}{2}\left( v_{12}\right)^2 + O(a)\ ,
\end{aligned}\eqn{termb}\eeq
with the field strength defined as $v_{mn} =
-i[D_m,D_n]$.

Note that the two  terms \Eq{terma} and \Eq{termb} each individually
violate the Euclidean rotation symmetry, but when added together they
yield
\beq
\frac{1}{g^2 a^2} \int d^2z\  \frac{1}{2}|D_m
s|^2-\frac{1}{2}[s^\dagger,s]^2 + \frac{1}{4} v_{mn} v_{mn} + O(a),
\eqn{sum}\eeq
with $s\equiv (s_1+is_2)/\sqrt{2}$.
This is identical to the bosonic action of the the target theory
\Eq{targ2}.  Note that even
though $\{x,y\}$ transform non-trivially under the discrete
$C_{2v}$ symmetry of the lattice in Fig.~\ref{fig:2dq4}, we see
that the leading term in our expansion in powers of $a$ is invariant
under rotations, with $v_m$ and $D_m$ transforming as  2-vectors and $s$ as a
scalar --- even though $s$ is constructed out of the real parts of $x$
and $y$, while the gauge fields  are constructed out of the
imaginary parts of $x$ and $y$.  Even better, the action \Eq{sum} exhibits an {\it internal}
 $U(1)$ symmetry consisting of phase rotations of $s$.  This symmetry is just the $U(1)_R$ symmetry of the target
theory; it does not exist as a symmetry in the lattice action.  This
bizarre and delightful behavior is characteristic of the supersymmetric
lattices I will present here.

To expand the fermionic part of the lattice action \Eq{lat2d} in powers of $a$,
I define the spinors
\begin{equation}
\psi= \begin{pmatrix} \lambda \cr \xi
 \end{pmatrix} \ ,\qquad 
\mybar\psi=  i\begin{pmatrix}\alpha & \beta \end{pmatrix}
\end{equation}
and the $\gamma$ matrices
\beq
\gamma_1=\sigma_3\ ,\quad \gamma_2=\sigma_1\ ,\quad \gamma_3 =
\sigma_2\ .
\eeq
After a little work one  finds that the fermionic part of the
lattice action \Eq{lat2d} reproduces the fermionic part of the target
theory's action in \Eq{targ2}, plus corrections of $O(a)$.  The
lattice theory has no fermion doubler modes, and the anomalous
$U(1)_R$ symmetry emerges in the fermion sector as well at leading
order in $a$, even though not present as a symmetry of the
lattice action. And, of course, we know that by yielding the $\CQ=4$
SYM theory in the continuum limit, somehow the full two dimensional
super-Poincar\'e symmetry has
emerged as well.

At least at the classical level then, one finds that the lattice
theory, expanded about the values \Eq{vevs} yields the desired target
theory in the limit 
\beq
a\to 0\ ,\quad g\to \infty\ ,\quad g_2 \equiv  a
g\ {\text{fixed}}.
\eeq
 At this point  I have provided no reason to expect
radiative corrections to respect the symmetries we have found at the
classical level.  For example, the two expressions in \Eqs{terma}{termb} indicate that bosonic
kinetic terms which do not respect two dimensional Lorentz invariance
are allowed by the $C_{2v}$ symmetry of the lattice, and that the
target theory only emerges if the two terms are added with precisely
related coefficients. One might expect that this delicate balance is
destroyed by quantum effects.  In fact, this does not happen, and that
is because the lattice action \Eq{lat2d} possesses an exact
supersymmetry that tames the radiative corrections.

\subsection{Exact lattice supersymmetry and radiative corrections}
\label{sec:3b}

In order to make explicit the supersymmetry of the lattice action
\Eq{lat2d}, it is convenient to develop a lattice superfield
notation, with the action of a supercharge $Q$ being translation along
a Grassmann coordinate $\theta$:  $Q=\frac{\partial\ }{\partial
  \theta}$.  We can define superfields constructed out of our lattice
variables (and a new auxiliary field $d$):
\beq
\begin{aligned}\hspace{.3in}
 \bfx_{\mathbf n} &=  x_{\mathbf n} +\sqrt{2}\,\theta \,\alpha_{\mathbf n}\\
    \bfy_{\mathbf n} &=  y_{\mathbf n}+\sqrt{2}\,\theta \,\beta_{\mathbf n}\\
    \boldsymbol{\Lambda}_{\mathbf n} &= \lambda_{\mathbf n} -
    \left(({\mybar x}_{\mathbf{n}-\xh} x_{\mathbf{n}-\xh} -
      x_{\mathbf n}{\mybar x}_{\mathbf n}) \right.\cr &\ \left.+({\mybar
        y}_{\mathbf{n}-\yh} y_{\mathbf{n}-\yh} -y_{\mathbf
        n} {\mybar y}_{\mathbf n})  +i d_{\mathbf n}\right)\,\theta\\
    \boldsymbol{\Xi}_{\mathbf n} &= \xi_{\mathbf n} + 2\left({\mybar
        x}_{\mathbf{n}+\xh} {\mybar y}_{\mathbf n}- {\mybar
        y}_{\mathbf{n}+\yh} {\mybar x}_{\mathbf n}\right)\,\theta\ .
  \end{aligned}
\eqn{sf2d}\eeq
The structure of these superfields is revealing.  We see that
application of the supercharge $Q=\frac{\partial\ }{\partial
  \theta}$ transforms $x$ into $\alpha$ and $y$ into $\beta$, as
one might expect from Fig.~\ref{fig:2dq4}.  However the site
fermion $\lambda$ transforms into a combination of the link bosons
$x$ and $y$,
which after the expansion about $x=y=\frac{1}{\sqrt{2}\, a}$ is
related to the hopping terms for the scalar $s$.  The diagonal
link fermion $\xi$ transforms into  a combination of the link bosons
related to the plaquette, which gives hopping terms for the gauge field
$v_m$.  In this  devious way  supersymmetry becomes entwined with translations
in the continuum limit of our lattice: the supercharge $Q$ does not
have a $\frac{\partial\ }{\partial
  x}$ in its definition, but the superfields it acts on are slightly nonlocal.
Note that in spite of this non-locality, the  supersymmetry transformations are all gauge
covariant, since the superfields I have written down are gauge
covariant.
\begin{figure}[thb]
\centerline{\epsfxsize=2.5 in \epsfbox{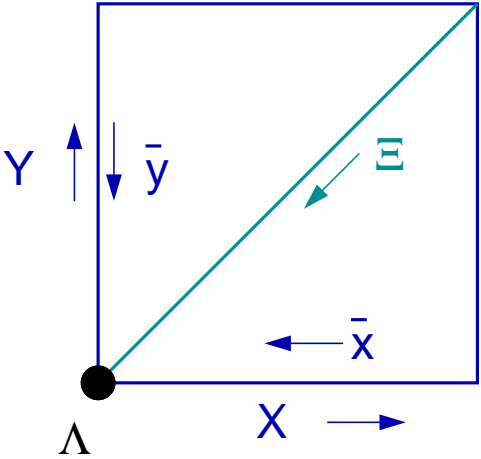}}
\caption{The lattice for SYM in $d=2$ with $\CQ=4$ supercharges, in
  terms of the superfields of \Eq{sf2d}}
\label{fig:2dq4SF}
\end{figure}

In terms of the superfields of \Eq{sf2d} the lattice may be drawn as in
Fig.~\ref{fig:2dq4SF}.  The lattice action \Eq{lat2d} may be
rewritten in terms of the superfields as 
\begin{equation}
\begin{aligned}
  S =& \frac{1}{g^2} \int\! d\theta\,\sum_{\mathbf n} \Tr\biggl[-\half
  \boldsymbol{\Lambda}_{\mathbf n}\,\partial_\theta
  \boldsymbol{\Lambda}_{\mathbf n}  \cr 
&
  -\boldsymbol{\Lambda}_{\mathbf n}(\mybar
  x_{\mathbf{n}-\xh}{\bfx}_{\mathbf{n}-\xh} - {\bfx}_{\mathbf
  n}\mybar x_{\mathbf n}\cr 
& \qquad
+\mybar y_{\mathbf{n}-\yh}
  {\bfy}_{\mathbf{n}-\yh}-{\bfy}_{\mathbf   n}\mybar y_{\mathbf n}
  ) \cr 
&
  - \,\boldsymbol{\Xi}_{\mathbf n }\,({\bfx}_{\mathbf
  n}{\bfy}_{\mathbf{n}+\xh}  -  {\bfy}_{\mathbf n 
  }{\bfx}_{\mathbf{n}+\yh})\biggr] 
  \eqn{sfact}
\end{aligned}
\end{equation}
This expression is not especially illuminating, except to show that
the supercharge $Q$ generates an exact symmetry, as the action
\Eq{sfact} is manifestly invariant under translations of $\theta$.

In order to analyze the stability of the theory under radiative
corrections, we can follow the following sequence of well-defined
steps:
\begin{enumerate}
\item We expand both  the superfields and the action about the point in
    moduli space $x=y=\frac{1}{\sqrt{2}\,a}$
in powers of $a$, keeping the form of the
    action manifestly supersymmetric;
\item By power counting we identify the general form
  of all local operators consistent with the exact symmetries of the
  lattice whose coefficients could receive divergent
  contributions.  At $\ell$-loops
  we determine how many powers of the coupling constant $g$ will accompany
  the operator,  make up the needed dimension with powers of
  the lattice spacing $a$, and consider the $a\to 0$ limit.
\item Each possible counterterm which violates the symmetries of
  the target theory and has a divergent coefficient presumably
  requires a
  fine-tuning in order to obtain the target theory in the $a\to 0$ limit. (From our
  classical analysis, we already know that all bad
  operators have vanishing coefficient in the  tree level action.)
\end{enumerate}
 As shown in ref.~\cite{Cohen:2003xe}, the two dimensional lattice being discussed here
has enough symmetry to forbid any supersymmetry violating divergences,
with the exception of a harmless correction to the vacuum energy. We
conclude that the target theory \Eq{targ2} is obtained in the 
continuum limit of our lattice theory without the need for fine-tuning.
 
(I should point out that the theories could be better behaved than I am
assuming. I have not considered the possibility of holomorphy
arguments which could protect the lattice against radiative
corrections even when there are no exact symmetries that will \cite{Terning:2003th}.)

\subsection{Infrared divergences and flat directions}
\label{sec:3c}

The second step in the above procedure requires elaboration.  The power
counting and dimensional analysis is done in a perturbative expansion
in the bare coupling $(g_d)^2=g^2 a^d$ of the $d$-dimensional target theory,
which is justified so long as that coupling is small and one does not
encounter infrared divergences.  Let us consider first the
applicability of perturbation theory. For SYM theories in $d=2$ and
$d=3$ 
dimensions, the gauge coupling $g_d$ is dimensionful and corresponds
to a fixed physical scale in the continuum limit.  Therefore in these
cases the bare coupling  in lattice 
units ($ (g_d)^2 a^{4-d}$) vanishes in the $a\to 0$
limit.  In $d=4$ dimensions $ (g_d)^2$ is dimensionless and will be
small as $a\to 0$ if the theory is either 
asymptotically free, or conformal and weakly coupled.  The SYM
theories in $d=4$ which can be analyzed using our methods 
 are $\CN=4$ (which is conformal in the continuum)
or $\CN=4$ broken by mass terms for the particles down to $\CN=2$ or
$\CN=1$ (which are asymptotically free theories).  Therefore the
power counting arguments are justified for each of these cases, with the
 exception of the strongly coupled $\CN=4$ theory, for which
there is no perturbative expansion in the bare coupling.  

As for infrared divergences encountered in a perturbative expansion of
the lattice theory, there are three possible sources.  In $d=4$
asymptotically free theories, there are the infrared divergences
of the sort familiar from QCD; they actually validate the perturbative expansion
by forcing us to take the bare coupling to zero in the continuum
limit. 

A second source of infrared divergences arises from the flat
directions of the lattice action.  In the two dimensional example
here, these correspond to the family of
transformations on the $x$ and $y$ lattice variables which leave the
action invariant. One class of such transformations are the gauge transformations, which form
a compact space and so are not a problem.  However others form a
noncompact space, such as shifting $x$ and $y$ everywhere on the
lattice by  the unit matrix
times two independent constants.  These latter global transformations
correspond to changing the size of the lattice spacing independently
in the $x$ and $y$ directions.  Obviously, the existence of such flat directions
implies that path integration over $x$ and $y$ will  diverge; small consolation that the exact lattice supersymmetry implies
that there are compensating fermion zeromodes!  Nevertheless, we can
treat the bosonic zeromodes exactly as in the case of
spontaneous breaking of a compact global symmetry: we can introduce a
small symmetry breaking term to the action, which we remove in the
large volume limit.  In the present case, such a term would take the
form 
\beq
\frac{1}{g^2} \sum_\bfn  \mu^2 a^2 \Tr\left[ \left(x_\bfn^2 - \frac{1}{2
    a^2}\right)^2 +
 \left(y_\bfn^2 - \frac{1}{2 a^2}\right)^2\right]
\eqn{sbt}\eeq
which serves to fix the values $x=y=\frac{1}{\sqrt{2}\, a}$ about
which we are expanding.  The corresponding fermionic zeromodes
(corresponding to the zero momentum $U(1)$ photinos of the target theory's
$U(k)$ gauge symmetry) may also be regulated or eliminated from the theory.

The third possible source of infrared divergences is special to
$d=2$. In dimensions $d>2$ the  symmetry breaking  terms \Eq{sbt} serve to fix
the moduli, and one can take $\mu\to 0$ in the infinite volume limit.
In  $d=2$ the situation is somewhat more complicated. There is no
spontaneous symmetry breaking in $d=2$ due to the log divergent
fluctuations of the would-be Goldstone bosons.  These fluctuations are
proportional to
$\ln(L/a)$, where $L$ and $a$ 
are the $IR$ and $UV$ cutoff lengths respectively.  In the present
case, if $\mu\to 0$ while $\mu L \sim 1$,  we can expect fluctuations
in $x$ (and $y$) to be given by $\delta x \sim  g_2 \ln (L/a)$.  So long
as we take the continuum and infinite volume limits to satisfy  $(a g_2)
\ln (L/a)\to 0$ as $a\to 0$, $L\to \infty$ and $g_2$ is held fixed,
then we are assured that $\delta x << \vev{x}$ and so our expansion
about $\vev{x} = \frac{1}{\sqrt{2}\, a}$ is justified, even in
$d=2$. What makes this case different than the usual analysis of
spontaneous symmetry breaking is that we are not trying to fix
$\vev{x}$ to equal a physical scale, but rather to equal the cutoff.

Of course the fluctuations  $\delta x $ may turn out to be much
smaller than $ g_2 \ln (L/a)$ due to non-perturbative dynamics; that is
a question of real physics.  If the $L$ dependence of fluctuations
does not go away, that probably signifies that the target theory
\Eq{targ2} is ill defined with a noncompact moduli space.

\section{SYM in $d=2$ with $\CQ=8$}
\label{sec:4}

As another example of a $d=2$ supersymmetric theory I now
briefly describe the lattice for $SYM$ in $d=2$ dimensions with
$\CQ=8$ supercharges.  The target theory consists of a $U(k)$ gauge field
$v_m$ with $m=1,2$; two Dirac spinors $\Psi_i$ with $i=1,2$; and four
real scalars $s_\mu$ with $\mu=0,\ldots,3$.  The Lagrangean of the
target theory is
\begin{equation}
\begin{aligned} 
  \CL&= \frac{1}{g_2^2} 
\Tr\Biggl[\frac{1}{4}
  v_{mn} v_{mn} +\frac{1}{2} (D_m s_\mu)^2 +\mybar\Psi_i \gamma_m D_m
  \Psi_i \\ &+ \mybar \Psi_i [s_0,\,\Psi_i] + i\mybar \Psi_i \gamma_3
  \tau^b_{ij}[{ s_b},\,\Psi_j]
  -\fourth[s_\mu,\,s_\nu]^2\Biggr]\ .
  \eqn{targ2b}  \end{aligned}
\end{equation}
In the above expression, $b=1,2,3$. This theory possesses an $SU(2)^3$
chiral symmetry (up to anomalies). An $SO(4)\simeq SU(2)\times SU(2)$
subgroup of this chiral
symmetry is explicit, with $\Psi_i$ transforming as a two-component doublet and
$s_\mu$ as a four-component vector; the remaining $SU(2)$ is somewhat
obscure and involves transformations between $\Psi_i$ and $\mybar
\Psi_i$, which is possible since all fields are in the (real) adjoint
representation of the gauge group.

\begin{figure}[t]
\centerline{\epsfxsize=2.55 in \epsfbox{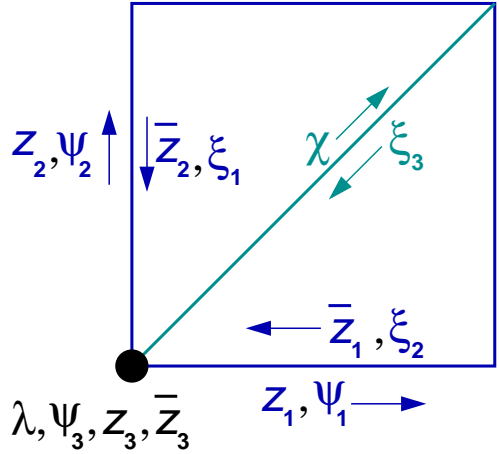}}
\caption{The lattice for SYM in $d=2$ with $\CQ=8$ supercharges. the
  $z_i$ are complex bosons, while the other fields are one-component fermions.}
\label{fig:2dq8}
\end{figure}

Our lattice for this theory respects two exact supercharges, with the
remaining six supersymmetries emerging only in the continuum limit. A
unit cell consists of three complex bosons $z_{1,2,3}$ and eight
one-component fermions $\psi_{1,2,3}$, $\xi_{1,2,3}$ $\chi$ and
$\lambda$.  The structure of the lattice is similar to the previous
example, and is shown in Fig.~\ref{fig:2dq8}.

The lattice action is given by
\begin{equation}\begin{aligned}
S= 
     \frac{1}{g^2}&\sum_{\bfn} \Tr\biggl[ 2\left\vert \epsilon_{ij}
       z_{i,\bfn} z_{j,\bfn + \ih} \right\vert^2 \cr &+ 2\left\vert z_{i,\bfn} z_{3,\bfn + \ih} -
        z_{3,\bfn} z_{i,\bfn} \right\vert^2 \cr &
\half \left(\mybar z_{i,\bfn
        -\ih} z_{i,\bfn-\ih} - z_{i,\bfn}\mybar z_{i,\bfn}+[\mybar
      z_{3,\bfn} ,\, z_{3,\bfn}]\right)^2 \\ &
     +\sqrt{2}\Bigl( \Delta_\bfn(\lambda,\mybar z_a, \psi_a)-
    \Delta_\bfn(\chi,\mybar z_a,\xi_a)
    \cr & \qquad\quad +\epsilon_{abc}\Delta_\bfn(\psi_a,z_b,\xi_c)\Bigr)
\eqn{d2lat}
  \end{aligned} 
\end{equation}
where the $\Delta$ functions have the same  structure  seen  in
 \Eq{delta}.


As before, analysis of the continuum limit is facilitated by making
the supersymmetry explicit.
A
superfield notation is again possible by introducing two independent Grassmann
variables $\theta$ and $\mybar \theta$.  There are seven bosonic
superfields $Z_i$, $\mybar Z_i$, and $S$, and three Grassmann
superfields $\Upsilon$, $\mybar \Upsilon$ and $\Xi$. 
Here are some examples of how several of the superfields are related to component fields:
 \begin{equation}
\begin{aligned}
{\bfz_{i,\bfn}} &= z_{i,\bfn} +\sqrt{2}\,\theta \psi_{i,\bfn} \cr &\quad
-\sqrt{2}\, \theta{\mybar\theta}
(\mybar z_{3,\bfn} z_{i,\bfn}- z_{i,\bfn} \mybar z_{3,\bfn + \ih})\ \\& \\
{\bfXi_\bfn} &= \xi_{3,\bfn} +\sqrt{2}\, \theta
\Bigl(G_\bfn-\sqrt{2}\,\epsilon_{ij}\,\mybar z_{i,\bfn+\jh}\mybar
z_{j,\bfn}
\Bigr) \cr & \quad
-\sqrt{2}\, \theta {\mybar\theta}
(\mybar z_{3,\bfn + \xh + \yh}\xi_{3,\bfn} -\mybar z_{3,\bfn}\xi_{3,\bfn}
)\ ,\\ & \\
%
{\bfU}_\bfn &=  \lambda_\bfn - 
\theta \Bigl(
\mybar z_{i,\bfn-\ih}  z_{i,\bfn-\ih} -
z_{i,\bfn}\mybar z_{i,\bfn} 
\cr &\quad +[\mybar{z}_{3,\bfn}, z_{3,\bfn}] +i d_\bfn\Bigr) 
-\sqrt2 \theta \mybar{\theta}[\mybar{z}_{3,\bfn}, \lambda_\bfn]\ .\\
\end{aligned}
\eqn{sfields2d}\end{equation}
 The action of
the two supercharges on the superfields is given by 
\begin{equation}
Q =\frac{\partial\  }{\partial \theta} + \sqrt{2}\,
\mybar\theta[\mybar z_3,\,\cdot \ ] \ ,\quad
\mybar Q =\frac{\partial\  }{\partial \mybar\theta} +   \sqrt{2}\,
\theta[\mybar z_3,\,\cdot \ ] \ ,\end{equation}
with the nontrivial anti-commutator
$\{Q,\mybar Q\} = 2\sqrt{2} [\mybar z_3,\,\cdot \ ]$.
To obtain an off-shell realization of the super-algebra required introducing
two auxiliary fields, $d$ and $G$.  Note that again the
supermultiplets are not entirely local, which is what leads to
supersymmetry charges being related to translations in the continuum.

 As in the $\CQ=4$ example,  the continuum
limit of the $d=2$, $\CQ=8$ lattice is defined as an expansion about the point in moduli space
\beq
\vev {z_1} = \vev{z_2} = \frac{1}{\sqrt{2}\, a}\ ,
\eeq
in the limit
\beq
a\to 0,\quad g\to \infty\ ,\quad g a = g_2\ {\text{fixed}}.
\eeq
Again one finds that at the classical level, the lattice theory
\Eq{d2lat} yields the target theory \Eq{targ2b} with the complex $z_3$ and the
real parts of $z_{1,2}$ forming the four real scalars $s_\mu$ of the
target theory, while the fermions arrange themselves as
\begin{alignat}{2}
  \mybar\Psi_1 &=
  \begin{pmatrix}
    -\chi & \psi_3
  \end{pmatrix} \ ,&\quad
  \mybar\Psi_2 &=
  \begin{pmatrix}
    \psi_1 & \psi_2
  \end{pmatrix} \ ,\\
 \Psi_1 &=
  \begin{pmatrix}
    \xi_1\\ \xi_2
  \end{pmatrix}\ , &\quad
  \Psi_2 &=
  \begin{pmatrix}
    \ \lambda\\ -\xi_3
  \end{pmatrix} \ ,\quad
  \eqn{ferms}
\end{alignat}

Furthermore, an analysis of the radiative corrections again reveals
that no fine-tuning is required to obtain the $\CQ=8$ target theory of
\Eq{targ2b} continuum limit.  I
refer you to Ref.~\cite{Cohen:2003qw} for details.

\section{Constructing the supersymmetric lattices}
\label{sec:5}

I purposely did  not  begin this talk by explaining  how to
construct the supersymmetric lattices I have presented.  The technique is described in
detail in 
Refs.~\cite{Kaplan:2002wv,Kaplan:2002zs,Cohen:2003xe,Cohen:2003qw},
and explaining it in a dark room tends to encourage people to go to
sleep. But at this point I have nothing to lose, so here is a brief
overview of the recipe.

For a target SYM theory in $d$ dimensions with a $U(k)$ gauge group
and $\CQ$ real supercharges:
\begin{itemize}
\item
Begin with a SYM theory in the continuum with $\CQ$ supercharges, but with the
much bigger gauge group $U({k} {N}^{ {d}})$;

\item
Reduce the theory to zero dimensions, yielding a  matrix model,
with  ${\CQ}$ supersymmetries, and a
$U( {k} {N}^{ {d}})\times G_R$ global symmetry

\begin{itemize}
\item {\sl For $ {\CQ}=4$, $G_R=SO(4)\times U(1)$;}
\item {\sl For $ {\CQ}=8$, $G_R=SO(6)\times SU(2)$;}
\item {\sl For $ {\CQ}=16$, $G_R=SO(10)$.}
\end{itemize}
 $G_R$ is an  ``$R$-symmetry''
 which does not commute with the supercharges. These $G_R$ groups are
 uniquely determined by the number of supercharges $\CQ$.

\item Identify a particular $(Z_{ {N}})^{ {d}}$ subgroup of the
  $U( {k} {N}^{ {d}})\times G_R$ symmetry.  Remove from the matrix model
  any variables which transform non-trivially under this discrete
  symmetry (a process called ``orbifolding'').
\end{itemize}
I will illustrate how this sequence of steps was followed to produce
the $d=2$, $\CQ=4$ lattice discussed in \S~\ref{sec:3}.

\smallskip
\noindent
{\bf Step 1:} We desire a matrix model with $\CQ=4$ supercharges
and a $U(kN^2)$ gauge symmetry.  This is easily constructed by
starting with the familiar $\CN=1$ SYM theory in four dimensions, and
then erasing the spacetime dependence of all the fields.  We are left
with the matrix model
\beq
S= \frac{1}{g^2}\Tr \left(\frac{1}{4} v_{mn}v_{mn} + \mybar \psi
  \mybar\sigma_m [v_m,\psi]\right)
\eqn{mmod}\eeq
where $m=0,\ldots,4$, $ \mybar\sigma_m=\{1,i\vec\sigma\}$ (where
$\vec\sigma$ are the three Pauli matrices), and  $v_{mn} =
i[v_m,v_n]$.  The fields $\psi$,
$\mybar \psi$ and $v_m$ are all $k N^2$ dimensional-matrices. The
symmetries of the action $S$ are: (i) $U(kN^2)$, under which
$v_m\to U v_m U^\dagger$, and similarly for $\psi$ and $\mybar \psi$;
(ii) an $R$-symmetry $G_R = SO(4)\times U(1)$ under which $v_m$
transforms as a neutral 4-vector, while $\psi$ and $\mybar \psi$
transform as charged spinors; and (iii) $\CQ=4$ supersymmetry,
consisting of the following transformations
\beq
\begin{aligned}
\delta\psi &= -iv_{mn} \sigma_{mn} \kappa\ ,\cr
\delta\mybar\psi &= iv_{mn}\mybar\kappa\mybar\sigma_{mn}\ ,\cr
\delta v_m &= -i\left(\mybar\psi\mybar\sigma_m\kappa
  -\mybar\kappa\sigma_m\psi\right)\ ,
\end{aligned}\eeq 
with $\mybar\sigma_m=\sigma_m^\dagger$ and $\sigma_{mn} =
\frac{1}{4}(\sigma_m\mybar\sigma_n-\sigma_n\mybar\sigma_m)$;
$\kappa$ and $\mybar\kappa$ are independent two-component Grassmann
spinor parameters.

\smallskip
\noindent
{\bf Step 2:} Next we project out a $Z_N\times Z_N$ symmetry, which
means that we identify a particular $Z_N\times Z_N$ subgroup of the
global $U(kN^2)\times SO(4)\times U(1)$ symmetry of our matrix model,
and then set to zero all fields except those which are neutral under
this discrete symmetry.  The result is to
render the $kN^2$-dimensional matrix variables sparse, inhabited by
only $N^2$ $k$-dimensional blocks.  These nonzero blocks will be
identified with $k$-dimensional matrix variables inhabiting the sites
or links of the $N^2$
cells of the two dimensional lattice, as shown in Fig.~\ref{fig:2dq4}. 

 The details of how this projection works are given
in Ref.~\cite{Cohen:2003xe} and won't be repeated here,  but the procedure may be
illustrated by  a toy example of two $3k$-dimensional matrix fields $\Phi_0$
and $\Phi_1$.  Assume that we have an action $S$  depending on
$\Phi_0$ and $\Phi_1$ which is invariant under a $U(3k)\times U(1)$
symmetry, where $\Phi_0$ and
$\Phi_1$ are both $U(3k)$ adjoints with $U(1)$ charges $0$ and $1$
respectively (the $U(1)$ plays the role of $G_R$).  We can
define a $Z_3\subset  U(3k)\times U(1)$ subgroup whose action on the
two fields is
\beq
\Phi_q\to e^{i(2 \pi q)/3}U \Phi_q U^\dagger,\quad U\equiv \begin{pmatrix}
  1 &&&\cr & \omega &&\cr &&\omega^2&  \end{pmatrix}
\ ,
\eeq
with $\omega=e^{2\pi i/3}\times \boldsymbol {1_k}$.  Setting to zero
all components of $\Phi_{0,1}$ except those which are invariant under
this $Z_3$ symmetry results in $\Phi_0$ consisting of three nonzero
$k$-dimensional blocks along the diagonal, and $\Phi_1$ consisting of three nonzero
$k$-dimensional blocks in the $\{12\}$, $\{23\}$ and $\{31\}$
positions.  We can interpret these remaining variables as 
degrees of freedom on a three-site, one-dimensional lattice with  
periodic boundary conditions---the three blocks in $\Phi_0$ become site
variables, while those in $\Phi_1$ become link variables.
Substituting the projected $\Phi$ matrices back into the original
action results in a lattice action that respects a $U(k)^3$ subgroup
of the original $U(3k)$ symmetry, with a $U(k)$ factor associated with
each of the three sites.  
The $Z_N\times Z_N$ projection of the matrix model \Eq{mmod} works
similarly, except that we get an $N$-site dimension for each of the
two $Z_N$ factors, resulting in a two-dimensional, $N^2$-site lattice.

The supersymmetric lattices  we construct suffer from two general
limitations: Firstly, the group $G_R$ needs to be able to contain a $Z_N^d$
subgroup, which means that it has to be at least of rank
$d$. Secondly, projecting out a $Z_N$ always breaks at least half of the
supersymmetries.  This means that lattices for
higher dimensional theories are in general less supersymmetric, and
therefore less well protected from radiative corrections. For example,
if we tried to construct a lattice action for $d=3$, $\CQ=4$ SYM
theory, we would find that the lattice would not respect any exact
supersymmetries, and so would presumably require fine-tuning.  It
is impossible to construct any lattice for $\CQ=4$ SYM in $d=4$
following our method, since for $\CQ=4$,   $G_R=SO(4)\times U(1)$ which is
only rank three.

\section{The bestiary of supersymmetric lattices}
\label{sec:6}

I now give a quick survey of the various supersymmetric lattices we have succeeded in
constructing.  The $\CQ=4$ and $\CQ=8$ lattices are discussed in
Refs.~\cite{Cohen:2003xe}, \cite{Cohen:2003qw} respectively;  a paper on  $\CQ=16$ target
theories is in preparation, and our work on Euclidean lattices in $d=1$ is unpublished.

\smallskip\noindent
{\bf One-dimensional lattices.}  We can construct one dimensional
lattices for supersymmetric quantum mechanics with $\CQ=2,4,8$, or $16$
supercharges, with $\CQ/2$ of those supercharges realized exactly on
the lattice.  No fine-tuning is required for any of these theories.
These may prove to be the most reasonable starting point for a numerical
investigation of supersymmetric lattices, as one may be able to use
conventional Hamiltonian techniques with which to compare the answers
obtained from a latticized path integral.  Furthermore, the $\CQ=16$
case  is interesting in its own right, as it is believed that the
large $N_c$ limit of such a theory is in fact $M$-theory \cite{Banks:1997vh}.
Understanding this theory could conceivably answer questions about
quantum gravity.

\begin{figure}[t]
\centerline{\epsfxsize=2.55 in \epsfbox{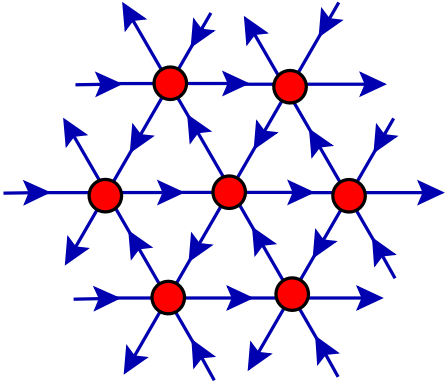}}
\caption{The lattice for SYM in $d=2$ with $\CQ=16$ supercharges.}
\label{fig:2dq16}
\end{figure}

\smallskip\noindent
{\bf Two-dimensional lattices.} Beside the $\CQ=4$ and $\CQ=8$
examples I have discussed, we can also construct the $d=2$, $\CQ=16$
lattice which has the triangular structure shown in
Fig.~\ref{fig:2dq16}. 
This lattice respects four exact
supersymmetries, and requires no fine-tuning.

\begin{figure}[t]
\centerline{\epsfxsize=2.55 in \epsfbox{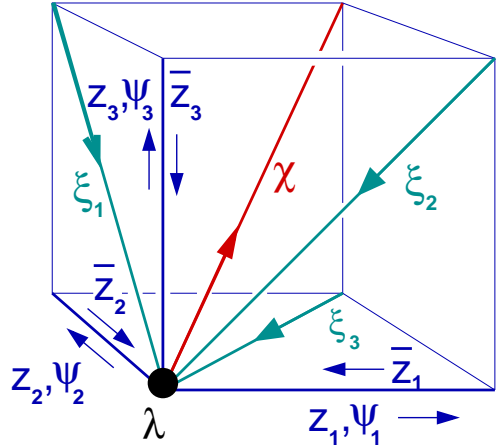}}
\caption{The lattice for SYM in $d=3$ with $\CQ=8$ supercharges.}
\label{fig:3dq8}
\end{figure}

\begin{figure}[t]
\centerline{\epsfxsize=2.55 in \epsfbox{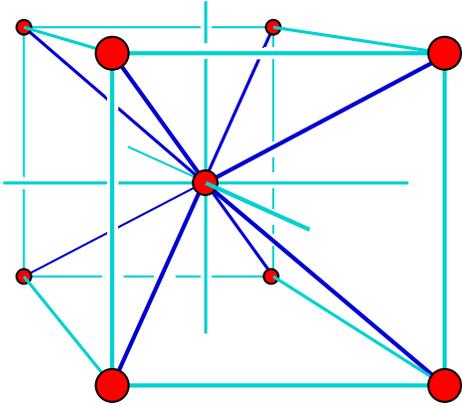}}
\caption{The bcc lattice for SYM in $d=3$ with $\CQ=16$
  supercharges. Link orientation and particle content are not shown.}
\label{fig:3dq16}
\end{figure}

\smallskip\noindent
{\bf Three-dimensional lattices.} We can construct $d=3$ lattices for
$\CQ=8$ and $\CQ=16$ which respect one and two exact supercharges
respectively.  The lattices take the form shown in Fig.~\ref{fig:3dq8}
and Fig.~\ref{fig:3dq16} respectively.  We find that two fine-tunings are indicated
in the $\CQ=8$ case, but none for $\CQ=16$. These theories both have
interesting features.  The $\CQ=8$ SYM theory possesses something
called ``mirror symmetry'', which has been analyzed in
Ref.~\cite{Kapustin:1999ha}, while the $\CQ=16$ theory is expected to
have a nontrivial infrared fixed point \cite{Seiberg:1998ax}.
required.

\smallskip\noindent
{\bf A four-dimensional lattice.} The only $d=4$ lattice we can
construct has as its target theory $\CN=4$ SYM.  The target theory is
expected to be finite, and  its large-$N_c$ limit is expected to have a
dual string theory description.  The lattice we construct is of the
$A_4^*$ type \cite{Conway:1991}; it respects a single exact
supersymmetry.  In this case it appears that the supersymmetry of the
lattice is not sufficiently powerful to eliminate the need for
fine-tuning, but we have not yet explored how many fine-tunings are

\smallskip\noindent
{\bf Spatial lattices.} Up to now I have only discussed Euclidean
lattices, but it is also possible to construct spatial lattices in
continuous Minkowski time for a Hamiltonian approach \cite{Kaplan:2002wv}.  The advantage
of these lattices is that, being lower by one dimension from the
corresponding Euclidean lattices, they possess twice as many exact
supersymmetries. This suffices, for example, to eliminate the need for
fine-tuning in the $d=3$, $\CQ=8$ case.   Fine-tuning may be required
for the $\CQ=16$ lattice in $d=3+1$ dimensions, but that is not
certain \cite{Kaplan:2002wv}.

\section{Future directions}
\label{sec:7}

I hope I have given you a sense of the  unusual and beautiful
properties of these supersymmetric lattices,  and how they
circumvent the apparent obstacles discussed in \S\ref{sec:2} ---
such as the need for chiral symmetries in the continuum --- in
 devious and novel ways.  I hope that further study will enlarge the
 class of supersymmetric theories that can be constructed on the
 lattice, perhaps making contact with the  ideas presented at this
 conference by Simon Catterall.  I am also optimistic that these
 lattices may prove useful for analytical studies of supersymmetry,
perhaps allowing  one to construct explicit Nicolai maps
\cite{Nicolai:1980nr} in a regulated theory, or leading to a better
understanding of duality in  SYM theories. Eventually I hope that such lattices could provide a
window onto quantum gravity through the connection discovered between
string theories and SYM in the large-$N_c$ limit.

The greatest obstacle now to numerical study of these theories is
the issue of simulating dynamical fermions.  Not only is there the
challenge of exactly massless fermions, but apparently there is a sign
problem as well \cite{Giedt:2003ve,Giedt:2003vy}.  It can be shown analytically that the sign
problem must disappear in the continuum limit for the $\CQ=4$ lattices, but
it is unknown what happens in the cases with more supersymmetry. In
any case, it would seem that analysis of one-dimensional lattices for
a path integral formulation of supersymmetric quantum mechanics would
be the technically most feasible place to start any
investigation.

\bibliography{latticeSUSY3}

\begin{thebibliography}{10}

\bibitem{Seiberg:1994aj}
N. Seiberg and E. Witten,
\newblock Nucl. Phys. B431 (1994) 484, hep-th/9408099,
\newblock 

\bibitem{Terning:2003th}
J. Terning,
\newblock (2003), hep-th/0306119,
\newblock 

\bibitem{Klebanov:2000me}
I.R. Klebanov,
\newblock (2000), hep-th/0009139,
\newblock 

\bibitem{Catterall:2001fr}
S. Catterall and S. Karamov,
\newblock Phys. Rev. D65 (2002) 094501, hep-lat/0108024,
\newblock 

\bibitem{Catterall:2003ae}
S. Catterall and S. Karamov,
\newblock Phys. Rev. D68 (2003) 014503, hep-lat/0305002,
\newblock 

\bibitem{Beccaria:2001tk}
M. Beccaria, M. Campostrini and A. Feo,
\newblock Nucl. Phys. Proc. Suppl. 106 (2002) 944, hep-lat/0110056,
\newblock 

\bibitem{Fujikawa:2002ic}
K. Fujikawa,
\newblock (2002), hep-th/0205095,
\newblock 

\bibitem{Montvay:2001aj}
I. Montvay,
\newblock (2001), hep-lat/0112007,
\newblock 

\bibitem{Feo:2002yi}
A. Feo,
\newblock (2002), hep-lat/0210015,
\newblock 

\bibitem{Feo:2003km}
A. Feo,
\newblock (2003), hep-lat/0305020,
\newblock 

\bibitem{Catterall:2003wd}
S. Catterall,
\newblock JHEP 05 (2003) 038, hep-lat/0301028,
\newblock 

\bibitem{Nishimura:2003tf}
J. Nishimura, S.J. Rey and F. Sugino,
\newblock JHEP 02 (2003) 032, hep-lat/0301025,
\newblock 

\bibitem{Itoh:2002nq}
K. Itoh et~al.,
\newblock JHEP 02 (2003) 033, hep-lat/0210049,
\newblock 

\bibitem{Kaplan:2002wv}
D.B. Kaplan, E. Katz and M. Unsal,
\newblock JHEP 05 (2003) 037, hep-lat/0206019,
\newblock 

\bibitem{Kaplan:2002zs}
D.B. Kaplan,
\newblock (2002), hep-lat/0208046,
\newblock 

\bibitem{Cohen:2003xe}
A.G. Cohen et~al.,
\newblock JHEP 08 (2003) 024, hep-lat/0302017,
\newblock 

\bibitem{Cohen:2003qw}
A.G. Cohen et~al.,
\newblock (2003), hep-lat/0307012,
\newblock 

\bibitem{Kaplan:1984sk}
D.B. Kaplan,
\newblock Phys. Lett. B136 (1984) 162,
\newblock 

\bibitem{Neuberger:1998bg}
H. Neuberger,
\newblock Phys. Rev. D57 (1998) 5417, hep-lat/9710089,
\newblock 

\bibitem{Kaplan:1999jn}
D.B. Kaplan and M. Schmaltz,
\newblock Chin. J. Phys. 38 (2000) 543, hep-lat/0002030,
\newblock 

\bibitem{Nishimura:1997vg}
J. Nishimura,
\newblock Phys. Lett. B406 (1997) 215, hep-lat/9701013,
\newblock 

\bibitem{Fleming:2000fa}
G.T. Fleming, J.B. Kogut and P.M. Vranas,
\newblock Phys. Rev. D64 (2001) 034510, hep-lat/0008009,
\newblock 

\bibitem{Banks:1982ut}
T. Banks and P. Windey,
\newblock Nucl. Phys. B198 (1982) 226,
\newblock 

\bibitem{Banks:1997vh}
T. Banks et~al.,
\newblock Phys. Rev. D55 (1997) 5112, hep-th/9610043,
\newblock 

\bibitem{Kapustin:1999ha}
A. Kapustin and M.J. Strassler,
\newblock JHEP 04 (1999) 021, hep-th/9902033,
\newblock 

\bibitem{Seiberg:1998ax}
N. Seiberg,
\newblock Nucl. Phys. Proc. Suppl. 67 (1998) 158, hep-th/9705117,
\newblock 

\bibitem{Conway:1991}
J.H. Conway and N.J.A. Sloane,
\newblock New York, USA: Springer-Verlag (1991) 703 P. (Grundlehren der
  mathematischen Wissenschaften 290).

\bibitem{Nicolai:1980nr}
H. Nicolai,
\newblock Phys. Lett. B89 (1980) 341,
\newblock 

\bibitem{Giedt:2003ve}
J. Giedt,
\newblock Nucl. Phys. B668 (2003) 138, hep-lat/0304006,
\newblock 

\bibitem{Giedt:2003vy}
J. Giedt,
\newblock (2003), hep-lat/0307024,
\newblock 

\end{thebibliography}
\bibliographystyle{h-elsevier2.bst} 


\end{document}